# Spectra and Light Curves of GRB Afterglows

A. G. Tolstov[1] and S. I. Blinnikov[1,2*]

[1]*Institute for Theoretical and Experimental Physics,
ul. Bol'shaya Cheremushkinskaya 25, Moscow, 117259 Russia*

[2]*Sternberg Astronomical Institute, Universitetskii pr. 13, Moscow, 119992 Russia*

**Abstract**—We performed accurate numerical calculations of angle-, time-, and frequency-dependent radiative transfer for the relativistic motion of matter in gamma-ray burst (GRB) models. Our technique for solving the transfer equation, which is based on the method of characteristics, can be applied to the motion of matter with a Lorentz factor up to 1000. The effect of synchrotron self-absorption is taken into account. We computed the spectra and light curves from electrons with a power-law energy distribution in an expanding relativistic shock and compare them with available analytic estimates. The behavior of the optical afterglows from GRB 990510 and GRB 000301c is discussed qualitatively.

Key words: *plasma astrophysics, hydrodynamics, and shock waves; gamma-ray bursts.*

## INTRODUCTION

The nature of the central sources of cosmic gamma-ray bursts (GRBs) has not yet been established. However, it is clear that GRBs with afterglows are at cosmological distances and release energy $\sim 10^{51}$ erg on a time scale of the order of 100 s. The observed GRB peculiarities (nonthermal spectra, rapid temporal variability) require an ultrarelativistic motion of emitting plasma with characteristic Lorentz factors $\Gamma \sim 100-300$ (see Piran 2000; Blinnikov 2000).

In the standard GRB model (Rees and Mészáros 1992), a photon–lepton fireball is produced (see Postnov 1998; Piran 2000). Initially, however, the GRB energy can also be electromagnetic (Usov 1994; Spruit 1999; Blandford 2002) and it probably propagates in a narrow cone (jet). The observed gamma-ray photons are generated by a nonthermal mechanism at the fronts of relativistic shocks (although the apparent nonthermal spectrum can also be explained in terms of the model of optically thick shells moving at relativistic velocities; see Blinnikov *et al.* 1999).

Here, we develop a technique for solving the angle-, time-, and frequency-dependent transfer equation, which is based on the method of characteristics. It can be applied to the motion of matter with a Lorentz factor up to 1000. The main object of application of this technique must be the early generation phases of gamma-ray emission (during collisions between internal shocks), for which the optical-depth effects can be noticeable. For now, however, we consider the radiation from the matter behind the front of an external shock and use an analytic solution (Blandford and McKee 1976) to describe the post-shock matter by taking into account the synchrotron self-absorption [cf. Downes *et al.* (2002), where the self-absorption was disregarded, but the hydrodynamics and spectrum of ultrarelativistic particles were computed in a self-consistent way]. We computed the spectra and light curves from electrons with a power-law energy distribution in an expanding relativistic shock and compare them with available analytic estimates.

## FORMULATION OF THE PROBLEM

One of the most popular models for GRB afterglows involves the propagation of a relativistic shell being decelerated by an external medium. The relativistic shock heats up the captured matter as it enters the shell and causes the particles to be accelerated to ultrarelativistic energies. The X-ray and optical afterglows from GRBs in these models are associated with the nonthermal (synchrotron) radiation of relativistic particles at the front of an external shock being decelerated in a circumstellar or interstellar medium (Mészáros and Rees 1997). Consider this problem in more detail by highlighting the most important points.

[*]E-mail: sergei.blinnikov@itep.ru

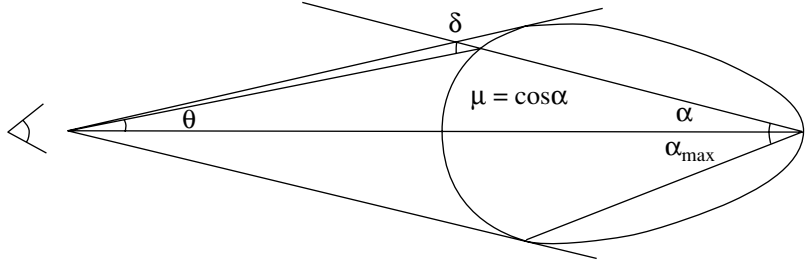

**Fig. 1.** The shape of the surface (the quasi-ellipsoid on the right) from which photons reach a remote observer (on the left) simultaneously. The explosion center is located at the vertex of the angle $\alpha$. The farthest point of the visible surface lies at a small distance of $\sim (1-\beta)$ of the semimajor axis from the explosion center; the semiminor axis is $\sim 1/\gamma$ of the semimajor axis.

### The Propagation of Radiation from a Relativistic Shell

Because of the high shock velocity, light from the ellipsoidal structure shown in Fig. 1 reaches the observer at a certain time. Let us determine the shape of the surface more accurately.

Consider an emitting spherical shell of initial radius $R(t_0) = R_0$ with an observer located at distance $D$ from its center. The shell begins to expand as $R = R(t)$. We assume that the time $t_0$ at which the shell expansion begins corresponds to the time $t_{0\text{obs}}$ at which the observation begins; $t_{0\text{obs}} = t_0 + (D - R_0)/c$, where $c$ is the speed of light.

The radiation from points of a sphere with a radius depending on the cosine of the angle $\mu = \cos\alpha$ will reach the observer at some time $t_{\text{obs}}$. For convenience, $t_{\text{obs}}$ is defined in such a way that it is equal to zero at the arrival time of the first signal on the shell motion. To determine the shape of the surface from which the radiation arrives, we take into account the fact that the time at which the propagation of photons from points of a sphere with a $\mu$-dependent radius $R$ begins is the same and is specified only by $t_{\text{obs}}$.

In other words,

$$t + \frac{(D^2 + R^2 - 2RD\mu)^{1/2}}{c} = t_{\text{obs}} + t_{0\text{obs}}.$$

The surface shape can be determined from this equation by substituting in $t = R^{-1}(t)$. If we consider the simplest shell propagation equation

$$R = R_0 + \beta c(t - t_0), \quad 0 < \beta < 1,$$

where $\beta$ is the $v/c$ ratio and $\gamma = (1-\beta^2)^{-1/2}$ is the Lorentz factor, then we obtain the equation of the surface

$$\frac{R - R_0}{\beta c} + t_0 + \frac{(D^2 + R^2 - 2RD\mu)^{1/2}}{c} = t_{\text{obs}} + t_{0\text{obs}},$$

$$R = \frac{\beta c t_{\text{obs}} + R_0(1-\beta)}{1 - \mu\beta} \quad (D \gg R).$$

As we see, in the approximation $D \gg R$, this equation is the equation of an ellipse (Rees 1967). In Fig. 1, the shape is more complex, because it corresponds to a variable velocity, as suggested by the solution of Blandford and McKee (1976).

Each point on the sphere is characterized by the intensity $I_0(\mu, r, \nu_0, \cos\delta_0)$ in the intrinsic frame of reference. In the observer's frame of reference, we write this intensity as $I(\mu, r, \nu, \cos\delta)$

The intensity along the line of light propagation does not change in the absence of emission and absorption sources and at the point of observation it will be the same as that at the point of radiation. Therefore, denoting $\mu' = \cos\theta$, we have for the flux

$$F_\nu = 2\pi \int\limits_{\cos\theta_{\max}}^{1} I(\mu, r, \nu, \cos\delta)\mu' d\mu'.$$

Here, it is more convenient to pass from integration over $\theta$ to integration over $\alpha$ (Fig. 1):

$$F_\nu = 2\pi \int\limits_{\mu_{\min}}^{1} I(\mu, r, \nu, \cos\delta)\mu'(\mu) d\mu'(\mu).$$

Denote $R(\mu)/D = p(\mu)$ and express $\cos\theta$ and $d\cos\theta$ in terms of $p(\mu)$ and $\mu$. The dependence $R(\mu)$ appears in the observer's frame of reference. It should be remembered that $p$ also depends on $t$ and $t$, in turn, can be expressed in terms of $t_{\text{obs}}$. However, to simplify our formulas, we will omit this dependence. For our subsequent calculations, we will need the following geometrical relations between the angles:

$$\cos\theta = \frac{1 - \mu p(\mu)}{l(\mu)}, \quad d\cos\theta = p^2\frac{\mu - p(\mu)}{l^3(\mu)},$$

$$\cos\delta = \frac{\mu - p(\mu)}{l(\mu)}, \quad \cos\delta_0 = \frac{\cos\delta - \beta}{1 - \beta\cos\delta},$$

$$\frac{\nu}{\nu_0} = \frac{1}{\gamma(\mu)(1 - \cos\delta\beta(\mu))},$$

where $l(\mu) = (1 + p^2(\mu) - 2p(\mu)\mu)^{1/2}$. For the flux, we then have

$$F_\nu(t_{\rm obs}) = 2\pi \int_{\mu_{\rm min}}^{1} \frac{(\mu - p(\mu))(1 - \mu p(\mu))}{(1 + p^2(\mu) - 2p(\mu)\mu)^2} p^2 I_0 \quad (1)$$
$$\times \left(r(\mu), \nu\left(\frac{\nu_0}{\nu}\right), \cos\delta_0(\cos\delta)\right) \left(\frac{\nu}{\nu_0}\right)^3 d\mu.$$

When the flux is calculated, the condition imposed on the lower integration limit $\mu_{\rm min}$ is determined by the angle that corresponds to the maximum angular size of the shell from the point of observation:

$$p'_\mu(1-\mu^2) - p(\mu - p) = 0$$

Our subsequent calculations are associated with a specific expression for the intensity $I(r, \nu_0, \cos\delta_0)$ on the shell surface and a specific shell propagation law $R(t)$.

### The Transfer Equation

For the intensity on the surface of a relativistic emitting shell to be calculated, we must solve the transfer equation in a comoving frame of reference. This is Eq. (2.12) from Mihalas (1980):

$$\frac{\gamma}{c}(1+\beta\mu)\frac{\partial I(\mu,\nu)}{\partial t} + \gamma(\mu+\beta)\frac{\partial I(\mu,\nu)}{\partial r} \quad (2)$$
$$+ \gamma(1-\mu^2)\left[\frac{(1+\beta\mu)}{r} - \frac{\gamma^2}{c}(1+\beta\mu)\frac{\partial\beta}{\partial t}\right.$$
$$\left. - \gamma^2(\mu+\beta)\frac{\partial\beta}{\partial r}\right]\frac{\partial I(\mu,\nu)}{\partial\mu} - \gamma\left[\frac{\beta(1-\mu^2)}{r}\right.$$
$$+ \frac{\gamma^2}{c}(1+\beta\mu)\frac{\partial\beta}{\partial t} + \left. + \gamma^2\mu(\mu+\beta)\frac{\partial\beta}{\partial r}\right]\nu\frac{\partial I(\mu,\nu)}{\partial\nu}$$
$$+ 3\gamma\left[\frac{\beta(1-\mu^2)}{r} + \frac{\gamma^2\mu}{c}(1+\beta\mu)\frac{\partial\beta}{\partial t}\right.$$
$$\left. + \gamma^2\mu(\mu+\beta)\frac{\partial\beta}{\partial r}\right]I(\mu,\nu) = \eta(\nu) - \chi(\nu)I(\mu,\nu).$$

Here, $\eta$ is the emission coefficient and $\chi$ is the absorption coefficient; the subscript 0 was omitted, because all quantities refer to the comoving frame.

Our numerical method of solution is described in the next section.

### Hydrodynamics

The transfer equation (2) explicitly or implicitly (via $\eta$ and $\chi$) includes variables of the medium: its velocity, density, temperature, etc. For these variables to be determined, we must solve the system of hydrodynamic equations. In general, the transfer and hydrodynamic equations constitute a combined system of equations. In our problem, however, we solve the transfer equation separately from the hydrodynamic equations. At the same time, to determine the variables of the medium, we use a self-similar solution for a relativistic shock (with a Lorentz factor of the post-shock matter $\gamma \gg 1$) in the spherically symmetric case for an ultrarelativistic gas (Blandford and McKee 1976). Let us give the formulas of this solution that we will need below.

Taking the law of time variations in the shock-front Lorentz factor in the form $\Gamma^2 \propto t^{-m}$ and choosing, for convenience, the self-similar variable $\zeta = [1 + 2(m+1)\Gamma^2](1 - r/t)$, we derive the following expressions for the pressure, velocity, and density of the post-shock matter from the conditions at the shock front:

$$p = \frac{2}{3}w_1\Gamma^2 f(\zeta), \qquad \gamma^2 = \frac{1}{2}\Gamma^2 g(\zeta), \qquad (3)$$
$$n' = 2n_1\Gamma^2 h(\zeta).$$

Here, $w_1$ is the enthalpy of the pre-shock matter, $n_1$ is its density, $\Gamma$ is the shock-front Lorentz factor, and $n'$ is the density of the post-shock matter in the observer's frame of reference. The matter density in the intrinsic frame of reference is related to the latter by $n' = \gamma n$.

Substituting Eqs. (3) into the hydrodynamic equations yields a system of equations for $f(\zeta)$, $g(\zeta)$, and $h(\zeta)$ as a function of $m$. Consider the case $m = 3$, which corresponds to the conservation of total shell energy. The shock energy contained in the layer between the radii $R_0(t)$ and $R_1(t)$ is given by the expression

$$E(R_0, R_1, t) = \int_{R_0}^{R_1} 16\pi p\gamma^2 r^2 dr.$$

If we substitute in the solution for the functions $f(\zeta)$, $g(\zeta)$, and $h(\zeta)$ at $m = 3$:

$$f = \zeta^{-17/12}, \quad g = \zeta^{-1}, \quad h = \zeta^{-7/4},$$

then the total energy will be $E = 8\pi w_1 \times \times t^3 \Gamma^2/17$, which gives the proportionality constant between $\Gamma^2$ and $t^{-3}$.

### Synchrotron Radiation

An accurate calculation of the spectrum requires knowing not only the hydrodynamic quantities but also the electron energy spectrum and the magnetic-field strength.

We assume that the electrons have a power-law distribution and that their total energy behind the shock front accounts for $\epsilon_e$ of the internal energy:

$$N(\gamma) = K_0 \gamma^{-p}, \gamma \geq \gamma_{\rm min,0} = \frac{\epsilon_e e_0}{n_0 m_e c^2},$$

where $m_e$ is the electron rest mass and $K_0 = (p-1)n_0\gamma_{\min,0}^{p-1}$.

The magnetic field is parametrized by the quantity $\epsilon_B$, which is equal to the fraction of the internal energy contained in the magnetic field: $B^2 = 8\pi\epsilon_B e$. The magnetic field is randomly oriented and decreases with time due to the adiabatic shell expansion. Other assumptions about the magnetic-field evolution and orientation weakly affect the resulting spectrum (Granot 1999).

After the electrons have derived energy immediately behind the shock front, they begin to lose it through adiabatic cooling determined by the solution of Blandford and McKee (1976) and through synchrotron radiation. This process was described in more detail by Granot and Sari (2001). We present only the basic formulas for synchrotron radiation used in our calculations.

The spectral power of a single electron averaged over the pitch angle is

$$P(\omega) = \frac{3^{5/2}}{8\pi}\frac{P_{\text{sy}}}{\omega_0}F\Big(\frac{\omega}{\omega_c}\Big),$$

where

$$P_{\text{sy}} = \frac{1}{6\pi}\sigma_T c B^2(\gamma_e^2 - 1), \quad \omega_c = \frac{3\pi}{8}\frac{eB}{m_e c}\gamma_e^2$$

and $F(u)$ is the standard function of synchrotron radiation (Rybicki and Lightman 1979). The synchrotron absorption coefficient is specified by the formula

$$\chi = \frac{1}{8\pi m_e \nu^2}\int\limits_{\gamma_{\min}}^{\gamma_{\max}}d\gamma\,\frac{N(\gamma)}{\gamma^2}\frac{d}{d\gamma}\Big(\gamma^2 P(\omega,\gamma)\Big).$$

## NUMERICAL SOLUTION OF THE TRANSFER EQUATION

The numerical solution of the problem is based on the simple and well-known method of characteristics (Mihalas 1980). We consider the relativistic transfer equation (2) in the spherically symmetric case in a comoving frame of reference.

The main complexity of the equation is the presence of four independent variables. The linearity of the equation allows its complexity to be decreased by constructing the characteristics for a given velocity field along which the differential operator is a total differential. If we choose the rays by describing them by a set of parameters and define $s$ as some length along the ray, then we can determine the characteristics, the paths $[t(s), r(s), \mu(s), \nu(s)]$, in such a way that

$$\frac{dI}{ds} = \frac{dr}{ds}\frac{\partial I}{\partial r} + \frac{d\mu}{ds}\frac{\partial I}{\partial \mu} + \frac{d\nu}{ds}\frac{\partial I}{\partial \nu} + \frac{dt}{ds}\frac{\partial I}{\partial t}.$$

We then derive the following system of equations that describe the characteristics from Eq. (2):

$$\frac{dt}{ds} = \frac{\gamma}{c}(1 + \beta\mu),$$
$$\frac{dr}{ds} = \gamma(\mu + \beta),$$
$$\frac{d\mu}{ds} = \gamma(1-\mu^2)\Big[\frac{1+\beta\mu}{r} - \frac{\gamma^2}{c}(1+\beta\mu)\frac{\partial\beta}{\partial t}$$
$$- \gamma^2(\mu+\beta)\frac{\partial\beta}{\partial r}\Big],$$
$$\frac{d\nu}{ds} = \gamma\Big[\frac{\beta(1-\mu^2)}{r} + \frac{\gamma^2}{c}(1+\beta\mu)\frac{\partial\beta}{\partial t}$$
$$+ \gamma^2\mu(\mu+\beta)\frac{\partial\beta}{\partial r}\Big]\nu.$$

With the introduction of the characteristic rays, the transfer problem simplifies to

$$\frac{dI(s)}{ds} = \eta(s) - \chi'(s)I(s),$$

where

$$\chi'(s) = \chi(s) + 3\gamma\Big[\frac{\beta(1-\mu^2)}{r}$$
$$+ \frac{\gamma^2\mu}{c}(1+\beta\mu)\frac{\partial\beta}{\partial t} + \gamma^2\mu(\mu+\beta)\frac{\partial\beta}{\partial r}\Big]$$

The characteristics are numerically computed by the fourth-order Runge–Kutta method with an adaptive step. The step is adaptive, because at $\beta$ close to unity, a small increment along the ray can lead to a significant angular jump comparable to the angular size of the emitting region.

Note also that the variable quantities in the method must be of the order of unity, as follows from the constraint imposed on the cosine of the angle. Therefore, it is convenient to use a new system of units. We denote the physical and dimensionless (used in the program) quantities by the subscripts $r$ and $p$, respectively. So, let

$$r_r = Rr_p, \quad t_r = Tt_p, \quad m_r = Mm_p.$$

Denoting

$$c_r = Cc_p, \quad I_r = YI_p,$$
$$\eta_r = J\eta_p, \quad \chi_r = X\chi_p,$$

where $c$ is the speed of light and $R, T, C, Y, J,$ and $X$ are constants, we then derive the following expression for $C, Y, J,$ and $X$ in terms of $R, T,$ and $M$ after a simple work with dimensions:

$$C = RT^{-1}, \quad Y = MT^{-2},$$
$$J = MR^{-1}T^{-2}, \quad X = R^{-1}.$$

*The Analytic Solutions Used to Test the Numerical Method*

Below, we give some of the analytic solutions that we used to test the numerical method.

Let us first consider the time- and frequency-independent transfer equation (2):

$$\gamma(\mu+\beta)\frac{\partial I(r,\mu)}{\partial r} + \gamma(1-\mu^2)$$
$$\times \left[\frac{1+\mu\beta}{r} - \gamma^2(\mu+\beta)\frac{\partial \beta}{\partial r}\right]\frac{\partial I(r,\mu)}{\partial \mu}$$
$$+ 3\gamma\left[\frac{\beta(1-\mu^2)}{r} + \gamma^2\mu(\mu+\beta)\frac{\partial \beta}{\partial r}\right]I(r,\mu)$$
$$= \eta(r,\mu) - \chi(r,\mu)I(r,\mu).$$

For constant emission and absorption coefficients and $\beta = 0$, the characteristics have the shape of straight lines $r\sqrt{1-\mu^2} = p$ ($p$ is the parameter) and the equation has the analytic solution

$$I(r,\mu) = \frac{\eta}{\chi}\left[1 - \exp\left\{-\chi\left(\mu r + \sqrt{R^2 - r^2(1-\mu^2)}\right)\right\}\right].$$

If we now assume $\beta$ to be constant, then the shape of the characteristics changes: $p(1+\beta\mu) = r\sqrt{1-\mu^2}$, and the solution becomes slightly more complex,

$$I(s) = I(r,\mu)$$
$$= \frac{\eta}{\chi}\left\{1 - \exp\left[-\chi\left(\frac{\gamma\mu r}{1+\beta\mu} - \frac{\gamma\xi(p)R}{1+\beta\xi(p)}\right)\right]\right\},$$

where

$$\xi(p) = \frac{-\beta p^2 - \sqrt{\beta^2 p^4 + (R^2 + \beta p^2)(R^2 - p^2)}}{R^2 + \beta^2 p^2}.$$

In the absence of absorption and for a constant emission coefficient on the right-hand side of the transfer equation, the analytic solution for the central characteristic ($\mu = \pm 1$) is

$$I(0) = \frac{\eta R}{\gamma(1-\beta)}, \qquad I(R) = 2\eta R\gamma,$$

at $\beta = $ const and

$$I(0) = \eta R\left[2\frac{\ln(1+B)}{B} - 1\right],$$
$$I(R) = 2\eta R\left(\frac{1-B}{1+B}\right)^{3/2}\left[\frac{1}{B}\ln\frac{1+B}{1-B} - 1\right],$$

where $B = \beta(R)$, at $\beta \sim r$.

To test the temporal component, we used the problem of intensity change near the surface of an emitting static transparent spherical shell with time when the emission coefficient abruptly changes from $\eta_0$ to $\eta_1$. If we use the following notation: $p = $

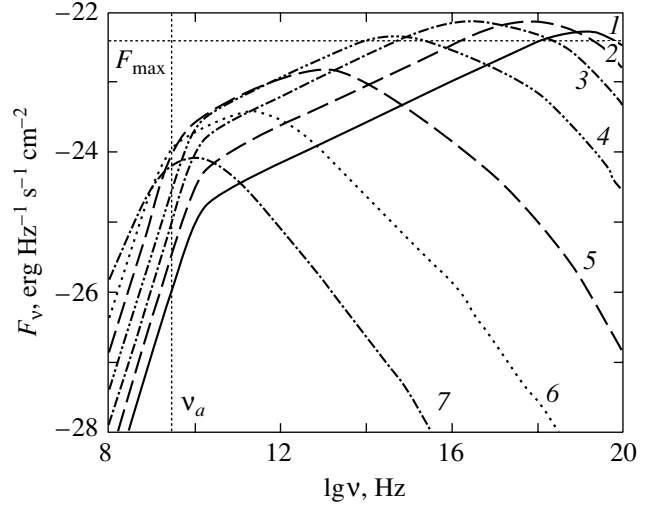

**Fig. 2.** Instantaneous afterglow spectra at various times $t = 10^N$ s, where $N$ is the number near the curve; $F_m$ and $\nu_a$ are the analytically estimated flux and self-absorption frequency, respectively.

$R/D$ is the ratio of the shell radius to the distance from the shell center to the point of observation, $t_+$ is the time measured from the beginning of the intensity change at the point of observation, $c$ is the speed of light, and $\tau = t_+c/D$, $d = 1 - p + \tau$, then the solution for this problem is

$$F_1(\tau) = G_1(\eta_1, (1-p)^2) - G_1(\eta_1, d^2)$$
$$+ G_1(\eta_0, d^2) - G_1(\eta_0, 1-p^2),$$

$$G_1(\eta, x) = \frac{\pi\eta R}{4p}\left[\frac{2}{3}\frac{(1-p^2)^3}{x^{3/2}} - 2\frac{(1-p^2)^2}{x^{1/2}}\right.$$
$$\left. + 2(1-p^2)x^{1/2} - \frac{2}{3}x^{3/2}.\right]$$

Here, $\tau$ changes from 0 to $\tau^* = \sqrt{1-p^2} - 1 + p$, i.e., within the interval during which the changes at the point of observation occur, and $F_1(\tau)$ is the flux. In the static case, this flux is

$$F = 2\pi\int_0^p \frac{(\mu-p)(1-\mu p)p^2}{(1+p^2-2p\mu)^2}I(\mu^*)\,d\mu,$$

where

$$\mu^* == \frac{\mu-p}{(1+p^2-2p\mu)^{1/2}}$$

is the cosine of the angle to the normal to the sphere surface on which the intensity depends.

Comparison of the solutions considered above with the calculations based on our numerical method shows that the numerical method is applicable to the motion of matter up to Lorentz factors $\gamma \sim 1000$ with an error less than 1%.

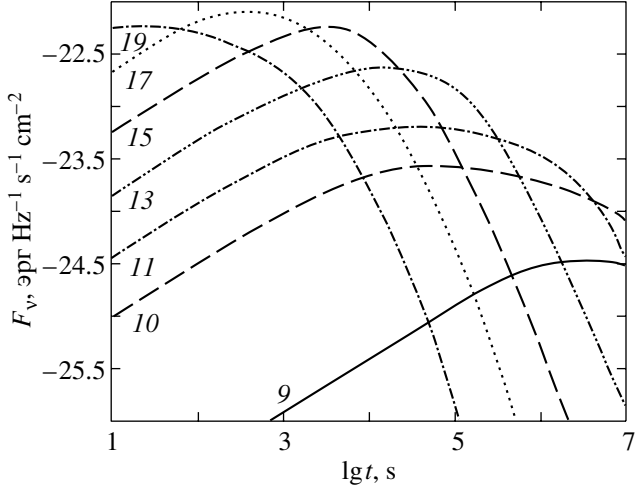

**Fig. 3.** Afterglow light curves for a set of frequencies $\nu = 10^N$ Hz, where $N$ is the number near the curve.

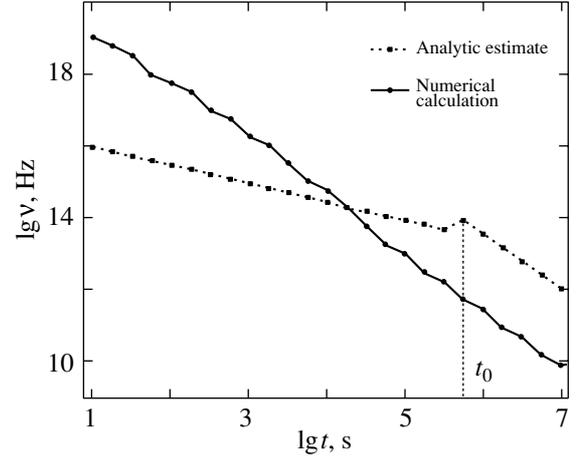

**Fig. 4.** Comparison of the frequencies that correspond to the maximum flux of the instantaneous spectra in the numerical calculations and analytic estimates.

## RESULTS OF THE NUMERICAL SOLUTION

We chose the following parameters for our numerical calculations of the afterglow spectra. These include the parameters that describe the hydrodynamics: the energy $E_0$ released through the process that leads to a GRB and the ambient density $n_1$; the parameters that describe the radiation: the fraction of the internal energy contained in the magnetic field $\epsilon_B$, the fraction of the internal energy transferred to electrons $\epsilon_e$, and the power-law index in the electron energy distribution $p$; and one more parameter: the photometric distance to the GRB $D$.

Our main calculation, whose results are presented here, is based on the following published parameters: $E_0 = 10^{53}$ erg, $n_1 = 1$ cm$^{-1}$, $\epsilon_e = 0.5$, $\epsilon_B = 0.1$, $p = 2.5$, and $D = 10^{27}$ cm.

The large amount of released energy $E_0$ is related to the spherical symmetry of the problem, while the observed GRBs can represent a jet with a solid angle $\Omega$. The total energy will then be lower by a factor of $\Omega/4\pi$.

The computed spectra and light curves are shown in Figs. 2 and 3. Here and below, the time is measured in the observer's frame of reference. Let us compare our results with available theoretical estimates (Hurley et al. 2002). In these estimates, the synchrotron spectrum is described by the maximum flux $F_{\max}$ and three characteristic frequencies $(\nu_{\min}, \nu_c, \nu_a)$, where $\nu_{\min}$ is the synchrotron frequency of the electron with minimum energy whose Lorentz factor is $\gamma_{\min,0}$, $\nu_c$ is the cooling frequency, and $\nu_a$ is the self-absorption frequency. For these four parameters, the theoretical estimates are given by the formulas

$$\nu_a = 2 \times 10^9 \text{Hz} E_{52}^{1/5} n_1^{3/5} \epsilon_e^{-1} \epsilon_B^{1/5} = 4 \times 10^9 \text{Hz},$$

$$\nu_{\text{cool}} = 9 \times 10^{12} \text{Hz} E_{52}^{-1/2} n_1^{-1} \epsilon_B^{-3/2} t_{\text{day}}^{-1/2}$$
$$= 2.66 \times 10^{16} \text{Hz} t_s^{-1/2},$$

$$\nu_{\min} = 5 \times 10^{15} \text{Hz} E_{52}^{1/2} \epsilon_e^2 \epsilon_B^{1/2} t_{\text{day}}^{-3/2}$$
$$= 3.80 \times 10^{22} \text{Hz} t_s^{-3/2},$$

$$F_{\max} = 20 \text{mJy} E_{52} n_1^{1/2} \epsilon_B^{1/2} d_{28}^{-2} = 6.32 \times 10^{-23} \text{mJy}.$$

Between these frequencies, the spectrum is a power law with indices $(2, 1.3, -1/2, -p/2)$ for $t <$

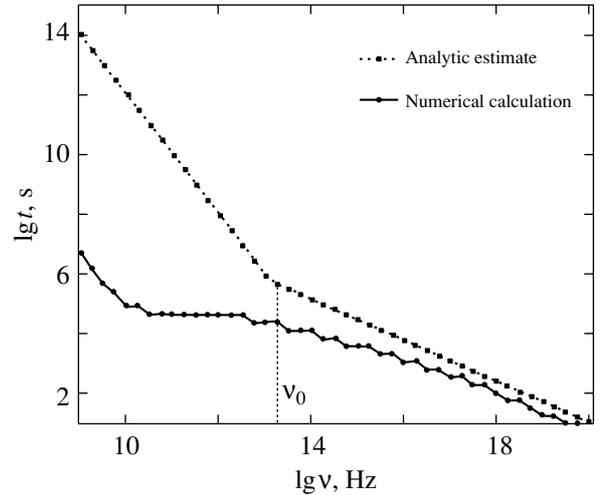

**Fig. 5.** Comparison of the times that correspond to the maximum flux of the light curves in the numerical calculations and analytic estimates.

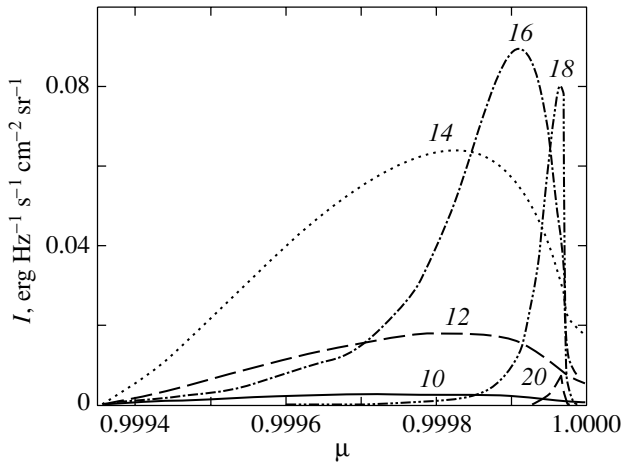

**Fig. 6.** The intensity on the radiation surface for a set of frequencies $\nu = 10^N$ Hz, where $N$ is the number near the curve.

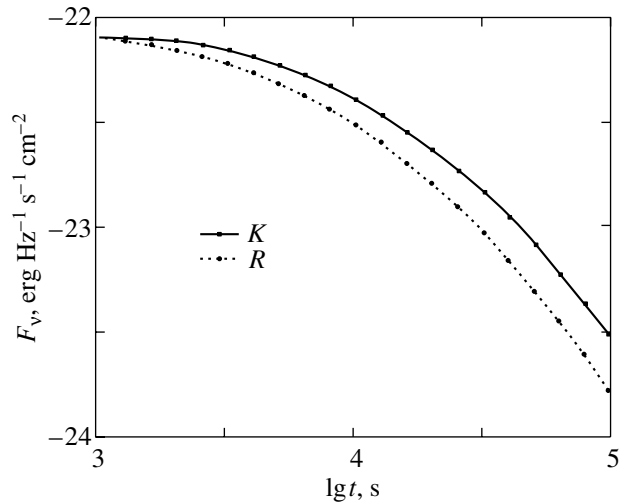

**Fig. 7.** Part of the theoretical light curve near the $R$ and $K$ bands.

$t_0 = 4.2 \times 10^5$ s and $(2, 1.3, -(p-1)/2, -p/2)$ for $t > t_0$.

Let us also compare the computed light curves with theoretical estimates (Sari *et al.* 1998), in which the characteristic times ($t_{\min}, t_{\text{cool}}$) were calculated from the flux $F_{\max}$ and the characteristic frequencies ($\nu_{\min}, \nu_{\text{cool}}, \nu_a$):

$$t_{\text{cool}} = 7.3 \times 10^{-6} E_{52}^{-1} n_1^{-2} \epsilon_B^{-3} \nu_{15}^2 \text{ day} = 63\nu_{15}^{-2} \text{ s},$$

$$t_{\min} = 0.69 E_{52}^{1/3} \epsilon_e^{4/3} \epsilon_B^{1/3} \nu_{15}^{-2/3} \text{ day}$$
$$= 2.37 \times 10^4 \nu_{15}^{-2/3} \text{ s}.$$

By introducing the frequency $\nu_0 = \nu_{\text{cool}}(t_0) = t_{\min}(t_0) = 1.14 \times 10^{13}$ Hz, we separate two cases: $t_{\min} < t_{\text{cool}}$ for $\nu > \nu_0$ and $t_{\min} > t_{\text{cool}}$ for $\nu < \nu_0$.

The results of our comparison are presented in Figs. 4 and 5. These figures show the frequencies for the spectra and the times for the light curves that correspond to the maximum flux calculated numerically and analytically, in accordance with the above estimates.

The plot of afterglow intensity versus observation angle ($\alpha \sim \theta$) (Fig. 6) at different frequencies reveals a bright ring attributable to the hotter matter at early afterglow stages. The higher the radiation frequency, the larger the contrast between the image center and edge. The same result was obtained by Granot *et al.* (1999).

The GRB energy must be released in a narrow cone, a jet. However, at early stages, the pattern for an observer near the cone axis will differ only slightly from the pattern produced by a spherical shock. At late stages, the jet becomes spherical.

In Fig. 7, the part of the theoretical light curve near the $R$ and $K$ bands is highlighted. We see that at longer wavelengths, the light curve passes to a decline more slowly than it does at shorter wavelengths. This chromatic behavior is characteristic of the optical afterglows from GRB 990510 (Stanek *et al.* 1999) and GRB 000301c (Jensen *et al.* 2001).

These objects deserve a more detailed study, but so far our model disregards several physical effects (the inverse Compton radiation, the Klein–Nishina effect, the non-power-law shape of the self-consistent electron spectrum, and others). Therefore, it cannot be directly used to interpret the spectra of early GRB afterglows. The work to take these effects into account is being continued.

## CONCLUSIONS

The most popular method for analytically estimating the spectra and light curves of GRB afterglows involves deriving the characteristic frequencies and times, determining their behavior, and calculating the corresponding flux and constructing the power-law segments of the spectra and light curves from the derived values. These methods are extensively presented in the literature (Granot and Sari 2001; Sari *et al.* 1998; Wijers and Galama 1999; Waxman 1997). The characteristic frequencies in different papers for the same cases occasionally differ by a factor of 70, including those in the same papers, suggesting that these treatments are insufficient.

Our calculations unambiguously describe the afterglow spectra and light curves in terms of the model under consideration by removing the uncertainty in characteristic parameters and without using the next approximations. The behavior of our results shows a relationship to the observed afterglows, which gives confidence that our technique can be used to study the early generation phases of gamma-ray emission (during collisions between internal shocks).

## ACKNOWLEDGMENTS

This study was supported by the Russian Foundation for Basic Research (project no. 02-02-16500). We are grateful to K.A. Postnov and D.K. Nadyozhin for valuable discussions.

*Translated by V. Astakhov*